\documentclass[superscriptaddress,twocolumn,prapplied,longbibliography]{revtex4-2}

\usepackage{graphicx}
\usepackage{dcolumn}
\usepackage{xcolor}
\usepackage{bm}
\usepackage[version=4]{mhchem}
\usepackage{amsmath}%

\begin{document}

\title{Nanoscale imaging of reduced forward bias at V-defects in green-emitting nitride LEDs}

\author{C. Fornos}
\email{camille.fornos@polytechnique.edu}
\affiliation{Laboratoire de physique de la matière condensée, CNRS, Ecole Polytechnique, IP Paris, 91128 Palaiseau, France}
 
\author{N. Alyabyeva}
\affiliation{Laboratoire de physique de la matière condensée, CNRS, Ecole Polytechnique, IP Paris, 91128 Palaiseau, France}

\author{W. Y. Ho}
\affiliation{Materials Department, University of California, Santa Barbara, California 93106, USA}

\author{C. Roubert}
\affiliation{Laboratoire de physique de la matière condensée, CNRS, Ecole Polytechnique, IP Paris, 91128 Palaiseau, France}

\author{T. Tak}
\affiliation{Materials Department, University of California, Santa Barbara, California 93106, USA}

\author{J. S. Speck}
\affiliation{Materials Department, University of California, Santa Barbara, California 93106, USA}

\author{C. Weisbuch}
\affiliation{Laboratoire de physique de la matière condensée, CNRS, Ecole Polytechnique, IP Paris, 91128 Palaiseau, France}
\affiliation{Materials Department, University of California, Santa Barbara, California 93106, USA}

\author{J. Peretti}
 \affiliation{Laboratoire de physique de la matière condensée, CNRS, Ecole Polytechnique, IP Paris, 91128 Palaiseau, France}

 \author{A. C. H. Rowe}
 \email{alistair.rowe@polytechnique.edu}
 \affiliation{Laboratoire de physique de la matière condensée, CNRS, Ecole Polytechnique, IP Paris, 91128 Palaiseau, France}

\date{\today}

\begin{abstract}
Record wall-plug efficiencies in long-wavelength, III-nitride light-emitting diodes (LEDs) have recently been achieved through improvements in electrical efficiency in devices containing V-defects. Numerical modeling suggests this may be due to reduced barrier heights for charge injection in thinned, low-Indium quantum wells parallel to semi-polar V-defect facets. To test this proposition, the tip of a scanning tunneling luminescence microscope is used as a local hole injector to map the optoelectronic properties of commercial, green-emitting LED heterostructures around V-defects with nanoscale spatial resolution. A 1 V reduction in the forward bias necessary for current injection at V-defect rims is observed. This, combined with the observation of small ($\approx$ 10 meV) blue shifts in the locally emitted electroluminescence, unambiguously confirms the charge injection mechanism.

\end{abstract}

\maketitle

\section{Introduction}
The wall-plug efficiency (WPE) of light-emitting diodes (LEDs) is defined as the output optical power per unit input electrical power. Today, blue-emitting, III-nitride LEDs exhibit peak WPEs exceeding 90 \% \cite{kuritzky2017}. However, as the Indium content in the InGaN quantum well alloys is increased to produce LEDs emitting at longer wavelengths, WPE drops precipitously \cite{weisbuch2015}, resulting in the so-called green gap \cite{usman2021}. There are on-going efforts in academic and industrial research laboratories to explore strategies aimed at improving WPEs of long-wavelength III-nitride LEDs, with the goal being to produce efficient white-light lamps based on color-mixed emitters. Success would have significant technological and economic advantages for lighting and micro-LED screen applications \cite{fan2023, hsiang2021}.

Recently, record-high WPE in green- and yellow-emitting nitride LEDs were obtained in devices containing high ($\approx 3 \times 10^8$ cm$^{-2}$) densities of V-defects emerging from threading dislocations \cite{jiang2019, zhang2020}, an improvement largely due to higher electrical efficiencies (EE). EE describes the conversion of the kinetic energy of injected charge, $q\textrm{V}_{\textrm{F}}$, where $\textrm{V}_{\textrm{F}}$ is the forward bias, into photon energy $h\nu$ according to \cite{kuritzky2017}: \begin{equation} \label{ee_def} \textrm{EE} = \frac{h\nu}{q\textrm{V}_\textrm{F}}. \end{equation} The inclusion of V-defects improves EE in green-emitting LEDs from $\approx$ 60 \% to more than 90 \% through a reduction in the forward bias of $\approx$ 1 V \cite{lheureux2020}. This improves WPE according to \begin{equation} \label{wpe} \textrm{WPE} = \textrm{IQE}\times \textrm{LEE}\times \textrm{EE},  \end{equation} where the light extraction efficiency (LEE) describes the ability of photons to escape the high-refractive index structure, and IQE is the internal quantum efficiency.

\begin{figure}
\includegraphics[width=1\columnwidth]{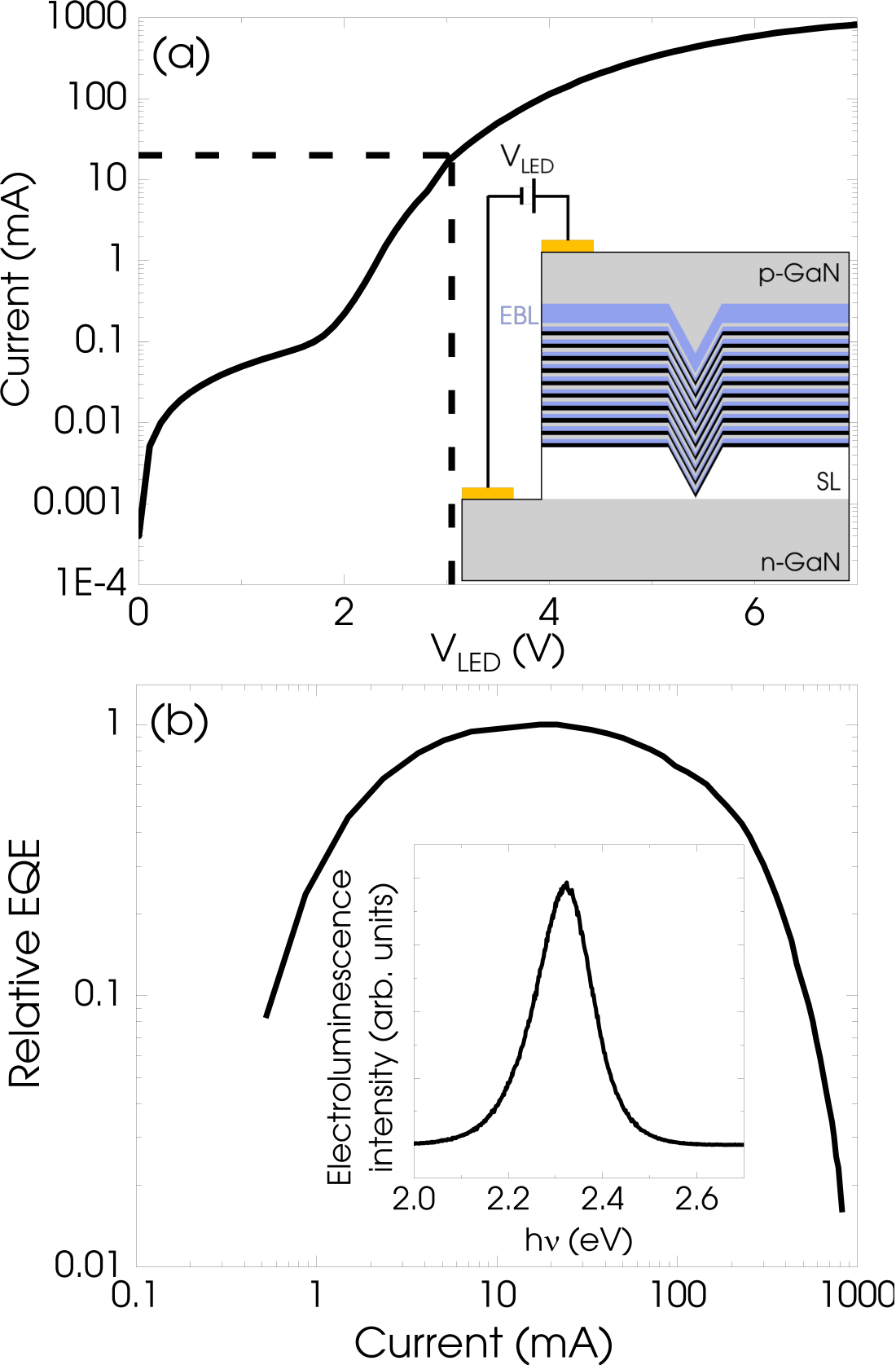}
\caption{\label{macro} Macroscopic characteristics of the functional LED. (a) The current-voltage characteristic at 300 K with a sketch of the LED heterostructure shown inset. (b) The relative EQE, along with the corresponding electroluminescence spectra measured at a forward current of 20 mA obtained when $\textrm{V}_{\textrm{LED}} \approx 3$ V as indicated by the dotted lines in (a).}
\end{figure}

The mechanism which yields this significantly improved EE is unclear. While early, ground-breaking work \cite{hangleiter2005} identified V-defects as being partially responsible (along with short $\left(0001\right)$-plane diffusion lengths \cite{rosner1999}) for high IQE despite the presence of threading dislocations \cite{hao2006, takahashi2000, han2013, yapparov2020}, a new element in the discussion of long-wavelength LEDs \cite{quan2014, li2016, ho2022, quevedo2024} proposes that V-defects act on EE by reducing internal potential barriers encountered by injected charge. This is thought to occur through a combination of the semi-polar nature of V-defect $\left\{ 10\overline{1}1 \right\}$ facets, along with the reduced widths and Indium contents of quantum wells oriented along these facets. As yet, no experimental work directly provides evidence for this \textit{charge injection} hypothesis.

An extensive experimental literature has progressively revealed the optoelectronic properties of V-defects \cite{hitzel2005, netzel2007, meyer2011, kim2014, sheen2015, zhou2017, ajia2018, yapparov2020}. Recent works employing far- \cite{chow2023} and near-field \cite{yapparov2024, yapparov2024b} imaging of the photo- and electro-luminescence suggest lateral hole transport following injection at V-defect facets prior to radiative recombination in $\left(0001\right)$-plane quantum wells. This ingredient is present in numerical modeling that predicts V-defect-induced improvements to EE \cite{ho2022, quan2014, li2016}, and was invoked to explain the characteristics of operational LEDs containing V-defects \cite{wu2016}. However, these approaches do not reveal the underlying physical mechanism for improved EE since this requires local optoelectronic measurements at length scales smaller than typical V-defect sizes ($\approx$ 100 nm). Two works have made attempts in this direction. The first employs high spatial resolution, self-emissive electron microscopy to provide local electrical information without correlating it with local optical properties \cite{tak2023}, and the second compares separate conductive atomic force microscopy measurements with near-field luminescence imaging \cite{li2019c}, but only with spatial resolutions that are, at best, comparable to V-defect sizes.

Here, we map the local optoelectronic properties of V-defects in a commercial, green-emitting LED heterostructures with spatial resolutions ($<$ 10 nm) by using the tip of a scanning tunneling luminescence microscope (STLM) as a local hole injector. We provide the first direct experimental evidence to support the charge injection hypothesis.

\section{Device details}
The LED studied here is a commercial, green-emitting multi-quantum-well heterostructure manufactured by Seoul Viosys with a V-defect density of $\approx 5\times10^8$ cm$^{-2}$. As indicated in the cross-sectional sketch inset in Fig. \ref{macro}(a), the n-GaN neutral zone of the LED is first grown on a substrate (here, patterned sapphire). A superlattice (SL) is used to seed the growth of V-defects (one of which is sketched), and the following multi-quantum well heterostructure consists of 10, 13 nm-thick repeats of a AlGaN/GaN/InGaN tri-layer (purple, gray, and black respectively in the inset of Fig. \ref{macro}(a)). This is the LED's space charge zone. A 20 nm-thick AlGaN electron blocking layer (EBL) is then grown, and the heterostructure is terminated by a 60 nm-thick p/p$^{++}$-GaN layer which forms the LED's p-type neutral zone. The p-GaN layer fills the V-defects as shown, and the heterostructure's surface is smooth. Indeed, atomic steps are clearly visible in scanning tunneling images of the surface at 300 K as discussed in Appendix \ref{appendix_STLM}. 

The LED is contacted with concentric, circular contacts (see photograph in Appendix \ref{appendix_LED}, Fig. \ref{photo_LED}) as indicated by the yellow rectangles in the cross-sectional sketch shown inset in Fig. \ref{macro}(a), and is biased using a voltage source, $\textrm{V}_{\textrm{LED}}$, to obtain a 300 K current-voltage characteristic (see Fig. \ref{macro}(a)). As shown inset in Fig. \ref{macro}(b), in forward bias ($\textrm{V}_{\textrm{LED}} > 0$) the LED emits light in the green part of the spectrum, with a centroid at $h\nu =$ 2.3 eV. The absolute \textit{external} quantum efficiency (EQE) of the LED is not directly measured here because the collection efficiency of the experimental optics is not known. The \textit{relative} EQE, calculated as the measured light intensity divided by the forward bias current, is however shown in Fig. \ref{macro}(b). The usual current dependence is observed, with a low-current regime limited by Shockley-Read-Hall recombination, and a high-current (or droop) regime limited by Auger-Meitner recombination \cite{iveland2013}. At intermediate currents, the relative EQE is independent of current as expected in the radiative limit. It is maximized at an injected current of 20 mA obtained when $\textrm{V}_{\textrm{LED}} = 3$ V as seen by the dotted lines in Fig. \ref{macro}(a). This particular value of $\textrm{V}_{\textrm{LED}}$ is taken to be the LED's macroscopic forward bias, $\textrm{V}_{\textrm{F}}$. This information can be used to calculate EE = 77 \% according to Eq. (\ref{ee_def}), which is the current state-of-the-art for green-emitting III-nitride LEDs (see Appendix \ref{appendix_LED} for further discussion).

\begin{figure*}
\includegraphics[width=2\columnwidth]{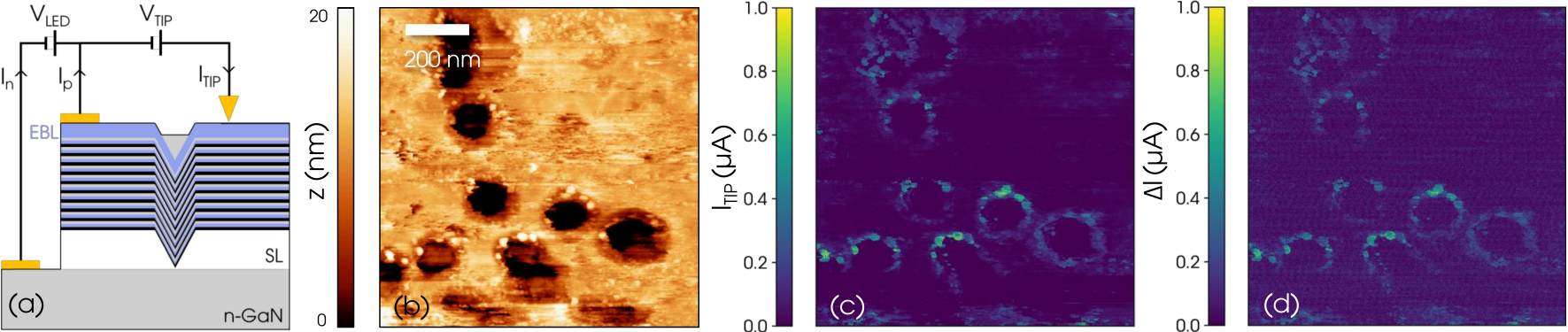}
\caption{\label{experiment} (a) A schematic drawing showing the addition of the STLM scanning tip used to locally inject holes ($\textrm{V}_{\textrm{TIP}} > 0$) into the ALE-treated heterostructure where the removal of the p-GaN exposes the V-defects at the sample surface as seen in the topography, (b). The topography is obtained using V$_{\textrm{TIP}} = 5.5$ V and V$_{\textrm{LED}} = 0$ V with a set point current of $\textrm{I}_{\textrm{TIP}} = 2$ nA. (c) The I$_{\textrm{TIP}}$ map corresponding to the topography in (b) shows large excess current, up to the µA range, particularly near V-defect rims. (d) The $\Delta\textrm{I} = \textrm{I}_{\textrm{n}}-\textrm{I}_{\textrm{p}}$ map is identical to the I$_{\textrm{TIP}}$ map to less than 0.1 \% at the V-defect rims, demonstrating that tip-injected current passes through the heterostructure to the n-type ohmic contact.}
\end{figure*}

Since the V-defects are filled with the p-GaN layer they are not visible in the 300 K topography obtained using the STLM (see Appendix \ref{appendix_STLM} for further discussion). In order to locally study their optoelectronic properties with the scanning tip of the STLM, they must be exposed at the surface. To this end, the 60 nm-thick p-GaN layer at the top of the LED heterostructure was removed using atomic layer etching (ALE) \cite{ho2023}. As indicated in Fig. \ref{experiment}(a), the etch stops at the EBL of the $\left(0001\right)$-plane, but leaves the $\left\{10\bar{1}1\right\}$ facets of the V-defects themselves covered with remnant p-GaN. Only the rims of the V-defect EBL are exposed at the surface. The resulting heterostructure is contacted using the same mask as that used for the operational LED in Fig. \ref{macro}(a) (inset), but now the inner circular electrode makes a Schottky contact directly with the EBL (see Fig. \ref{experiment}(a)) and the LED is no longer functional. 

\begin{figure*}
\includegraphics[width=2\columnwidth]{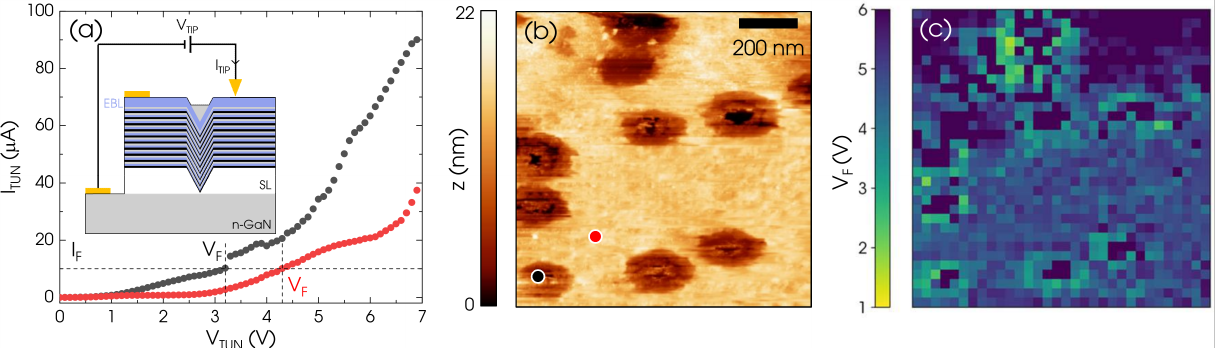}
\caption{\label{VF_maps} (a) Example current-voltage characteristics of the local LED (see inset). The black and red curves correspond to hole injection from the tip into a V-defect rim and a $\left(0001\right)$-plane at the points of the same color indicated in the topographic map, (b). (b) The topographic map obtained using $\textrm{V}_{\textrm{TIP}} = 7$ V and a current set point, $\textrm{I}_{\textrm{TIP}} = 0.5$ nA. (c) The simultaneous $\textrm{V}_{\textrm{F}}$ map (obtained for $\textrm{I}_{\textrm{F}}$ = 10 $\mu$A) showing a systematic reduction in $\textrm{V}_{\textrm{F}}$ at V-defect rims.}
\end{figure*}

\section{Local-probe opto-electronic measurements}
The typical STLM topography of the ALE surface at 300 K is shown in Fig. \ref{experiment}(b). In the schematic electrical circuit in Fig. \ref{experiment}(a), this image is obtained with $\textrm{V}_{\textrm{LED}} = 0$ V, $\textrm{V}_{\textrm{TIP}} = 6.5$ V. A set point current, I$_{\textrm{TIP}}$ = 10 nA, is used to stabilize the tip/surface distance in the usual STM manner. Note that since $\textrm{V}_{\textrm{TIP}} > 0$, the tip is injecting holes into the sample. Under these conditions the STM topography is of good quality, with the hexagonal bases of the inverted V-defect pyramids clearly visible. In the topography images, V-defect rims (brown), the remnant p-GaN filling the V-defects (black), and the $\left(0001\right)$-plane EBL surface (light brown) are distinguishable. The ALE treated surface no longer exhibits clean atomic steps like the original p-GaN surface of the LED (see Appendix \ref{appendix_STLM}, Fig. \ref{stm} for an example). Note also the typical height difference of about 20 nm between the remnant p-GaN and the EBL $\left(0001\right)$-plane surfaces.

Since the tip is injecting holes into a 2-terminal device, for the following discussion it is important to establish the path they take, either through the Schottky contact on the $\left(0001\right)$-plane EBL surface, thereby contributing to $\textrm{I}_\textrm{p}$ in the schematic drawing in Fig. \ref{experiment}(a), or through the heterostructure and out through the Ohmic contact on the n-type neutral zone of the LED, thereby contributing to $\textrm{I}_\textrm{n}$. A floating differential current measurement whose details are given in Appendix \ref{appendix_diff} allows for the difference, $\Delta\textrm{I} = \textrm{I}_\textrm{n}-\textrm{I}_\textrm{p}$, to be mapped simultaneously with both $\textrm{I}_\textrm{TIP} = \textrm{I}_\textrm{n}+\textrm{I}_\textrm{p}$ and the topography shown in Fig. \ref{experiment}(b). The $\textrm{I}_\textrm{TIP}$ map shown in Fig. \ref{experiment}(c) and the $\Delta\textrm{I}$ map in Fig. \ref{experiment}(d) are identical to less than 0.1 \% at the V-defect rims, and since $\Delta\textrm{I} > 0$, this indicates that at minimum 99.9 \% of $\textrm{I}_\textrm{TIP}$ flows through the multiple quantum wells of the heterostructure and out through the n-type neutral zone i.e. $\textrm{I}_\textrm{n} \gg \textrm{I}_\textrm{p}$. As will be seen below, the injection of $\textrm{I}_\textrm{TIP}$ on all parts of the surface except the remnant p-GaN filling the V-defects is accompanied by light emission from the quantum wells. 

A further key point to note is that currents well above the 10 nA set point are observed, particularly (but not only) when the tip is in the vicinity of the V-defect rims. In the current maps shown in Fig. \ref{experiment}, peak currents up to 1 $\mu$A are observed. The exact origin of this large excess current is unclear, but it is speculated that attractive image forces \cite{rideout1970} arising from unscreened charge in the semiconductor result in mechanical contact of the tip with the surface thereby yielding much larger currents. In this case, the tip-surface junction more closely resembles a local Schottky contact. This serendipitous effect allows for V$_{\textrm{TIP}}$ to be dropped entirely across the device (i.e. the local Schottky contact and the heterostructure), and removes the partial drop across the tunnel junction. Moreover, since $\textrm{I}_\textrm{n} \gg \textrm{I}_\textrm{p}$ and $\textrm{V}_{\textrm{LED}} = 0$ in the experiments reported here, the 3-terminal geometry shown in Fig. \ref{experiment}(a) can be approximately simplified to the 2-terminal geometry shown inset in Fig. \ref{VF_maps}(a). This is nothing more than a local LED formed by the macroscopic n-type neutral zone and the tip which acts as a local, non-Ohmic p-contact.

Inspired by scanning tunneling spectroscopy methods, the tip scan can be paused on a set of predefined points on a 32 $\times$ 32 square array and the current feedback loop opened. During this time V$_{\textrm{TIP}}$ is scanned from 0 V to 7 V in 0.1 V steps occurring every 10 ms while measuring the resulting current, $\textrm{I}_{\textrm{TIP}}$. In this way, local current-voltage characteristics of the 2-terminal device like those shown inset in Fig. \ref{VF_maps}(a) can be measured point-by-point. Between points in the square array, the feedback loop is closed and the tip is scanned to the next point to simultaneously obtain a topography map (like that shown in Fig. \ref{VF_maps}(b)) and electrical maps obtained from the local current-voltage characteristics. Two example characteristics are shown in Fig. \ref{VF_maps}(a), where the red curve corresponds to the tip position indicated by the red point in Fig. \ref{VF_maps}(b) (injection into the $\left(0001\right)$-plane EBL surface), and the black curve corresponds to the tip position indicated by the black point in Fig. \ref{VF_maps}(b) (injection into a V-defect rim).

Unlike the macroscopic current-voltage characteristic shown in Fig. \ref{macro}(a) where the LED's forward bias (V$_{\textrm{F}}$) is defined as being the value of V$_{\textrm{LED}}$ at which the relative EQE is maximized, in the case of the local characteristics measured under the STM tip, no spectral information about the local electroluminescence is available. The \textit{local} value of V$_{\textrm{F}}$ is therefore defined as the value of V$_{\textrm{TIP}}$ necessary to obtain an arbitrarily chosen value of I$_{\textrm{TIP}}$. In Fig. \ref{VF_maps}(a) this is indicated graphically for I$_{\textrm{TIP}} =$ 10 $\mu$A, as is the map of the resulting V$_{\textrm{F}}$ shown in Fig. \ref{VF_maps}(c). As will be shown, the conclusions drawn below are robust to this arbitrarily chosen value of the current. As Fig. \ref{VF_maps}(c) strikingly shows, local values of V$_{\textrm{F}}$ are systematically lower at V-defect rims than on the EBL's $\left(0001\right)$-plane surfaces. 

\begin{figure}
\includegraphics[width=1\columnwidth]{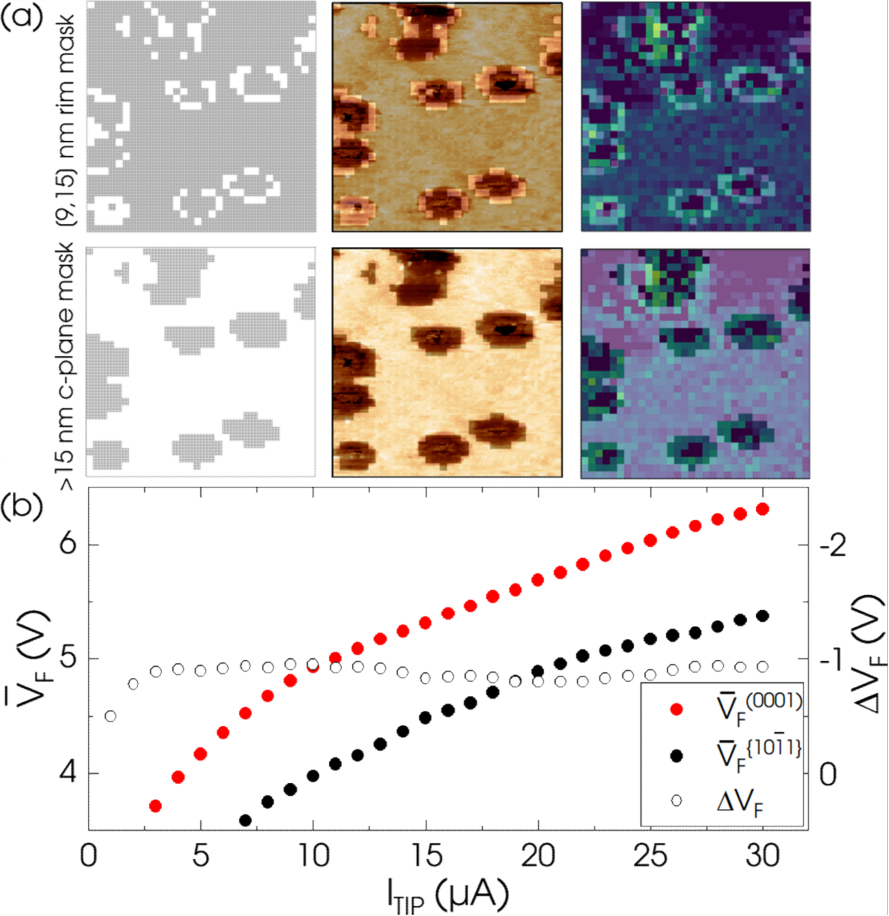}
\caption{\label{stats} (a) Masks defined to select image pixels corresponding to V-defect rims (top left) and the $\left(0001\right)$-plane (bottom left). The middle and right frames show these masks superimposed on the topography and on the ($\textrm{I}_\textrm{TIP} = 10 \mu$A) $\textrm{V}_\textrm{F}$ map respectively. (b) Average $\textrm{V}_\textrm{F}$ calculated for a range of arbitrarily chosen values of $\textrm{I}_\textrm{TIP}$ on pixels selected by the masks shown in (a). The difference ($\Delta\textrm{V}_{\textrm{F}}$), (black circles) is the measured difference in forward bias obtained when injecting holes directly into the V-defect rims rather than the heterostructure's $\left(0001\right)$-plane.}
\end{figure}

The correlation between local values of V$_{\textrm{F}}$ and topography can be emphasized further via a statistical treatment of the V$_{\textrm{F}}$ maps based on masks defined using the topography. The top frame of Fig. \ref{stats}(a) shows a rim mask which selects only pixels whose height in the topographic image in Fig. \ref{VF_maps}(b) lies between 9 and 15 nm, while the bottom-left $\left(0001\right)$-plane mask selects only pixels whose height in the topographic image is greater than 15 nm. This is most obvious when superposing the rim and $\left(0001\right)$-plane masks with the topography (middle frames). When applied to the V$_{\textrm{F}}$ maps (right frames), statistics on the selected pixels (shown partially whited out in both cases) can be made. More specifically, average forward bias values calculated over pixels filtered by the rim mask, $\bar{\textrm{V}}_{\textrm{F}}^{\left\{10\bar{1}1\right\}}$, and over pixels filtered by the $\left(0001\right)$-plane mask, $\bar{\textrm{V}}_{\textrm{F}}^{\left(0001\right)}$, can be obtained. Figure \ref{stats}(b) shows the result of this treatment.

Regardless of the arbitrarily chosen value of $\textrm{I}_\textrm{TIP}$, it can be seen that the average value of $\textrm{V}_\textrm{F}$ is always larger on the $\left(0001\right)$-plane (red points in Fig. \ref{stats}(b)) than at the V-defect rims (black points in Fig. \ref{stats}(b)). The absolute values of these quantities are not necessarily meaningful because the details of the local (Schottky) contact made between the tip and the EBL surface are unknown. However, the \textit{difference} between the two, $\Delta\textrm{V}_\textrm{F} = \bar{\textrm{V}}_{\textrm{F}}^{\left\{10\bar{1}1\right\}} - \bar{\textrm{V}}_{\textrm{F}}^{\left(0001\right)}$, is meaningful. As seen in Fig. \ref{stats}(b) (black circles), $\Delta\textrm{V}_\textrm{F} \approx -1 V$, independent of the arbitrarily chosen value of $\textrm{I}_\textrm{TIP}$. The only term in Eq. (\ref{wpe}) that is affected by changes in $\textrm{V}_\textrm{F}$ is the EE whose definition in Eq. (\ref{ee_def}) shows that the observed reduction in $\textrm{V}_\textrm{F}$ at V-defect rims corresponds to an increase in the local EE. 

\begin{figure*}
\includegraphics[width=2\columnwidth]{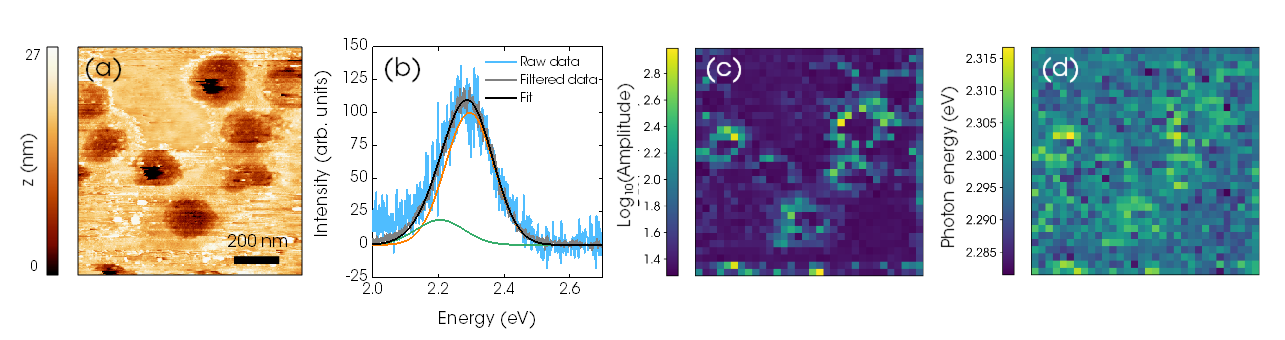}
\caption{\label{EL} (a) STM topography of the ALE heterostructure ($\textrm{V}_{\textrm{TIP}} = 6.5$ V, $\textrm{V}_{\textrm{LED}} = 0$ V, and $\textrm{I}_{\textrm{TIP}} = 0.5$ nA). (b) A typical local electroluminescence spectrum exhibiting a phonon replica (light green curve) and a centroid around 2.3 eV (orange curve).(c) The integrated electroluminesence intensity showing clear maxima around V-defect rims where excess current occurs (see also Fig. \ref{experiment}(c)). (d) Centroid energy map showing electroluminescence blueshifts on the scale of 10 meV near V-defect rims.}  
\end{figure*}

Since Eq. (\ref{ee_def}) also depends on the photon energy, it is useful to study the local electroluminescence spectrum resulting from the tip-injected hole current. This can be done by using the STLM in another mode where the tip scan is paused on a square array of 32 $\times$ 32 points. The hold time at each of these points is 10 s, during which the current feedback loop remains closed and the electroluminesence is collected after transmission through the substrate using an in-vacuum, steerable, aspherical lens (NA = 0.6). The collimated light is then transmitted through an ultra-high-vacuum viewport, and then coupled into a large-core (1.5 mm) fiber to a short focal length monochromator equipped with an electron-multiplied CCD camera. Fig. \ref{EL}(a) shows the topography obtained in one such experiment with an example electroluminescence spectrum shown in Fig. \ref{EL}(b). 

Typical spectra show features such as a phonon replica (light green curve in Fig. \ref{EL}(b)) and a centroid around 2.3 eV (orange curve in Fig. \ref{EL}(b)) close to that of the macroscopic electroluminescence spectrum obtained from the operational LED fabricated with this heterostructure (Fig. \ref{macro}(b), inset). The integrated electroluminescence intensity map is an injection map formed by measuring the far-field light intensity for each tip position. It is shown in Fig. \ref{EL}(c) and exhibits clear maxima around the V-defect rims due to the excess current which typically appears in these zones (Fig. \ref{experiment}(c)) for a given V$_{\textrm{TIP}}$. The centroid of the emission is the most important information for this discussion, an injection map of which is shown in Fig. \ref{EL}(d). Small but systematic blue shifts of the order of 10 meV are observed when the tip injects holes at the V-defect rims. This is much smaller than the linewidth of the macroscopic spectrum (Fig. \ref{experiment}(c)). It is also much smaller than the approximately 0.5 eV blue shifts observed in the photoluminescence of the V-defect facets relative to $\left(0001\right)$-planes (see Appendix \ref{appendix_PL}) \cite{yapparov2020, netzel2007, yapparov2024}. Such small blue shifts indicate that the majority, if not all, of the holes injected into V-defect rims are laterally transported to the $\left(0001\right)$-planes prior to radiative recombination with electrons in $\left(0001\right)$-plane quantum wells, a conclusion previously drawn from purely optical measurements \cite{meyer2011,yapparov2024b}. Since these are injection maps (i.e. far-field optical measurements of the electroluminescence made for a given tip position), the 10 meV blue shift in the electroluminescence may arise either from the small fraction of the injected holes which recombine directly in the V-defect facets, or from holes recombining in $\left(0001\right)$-plane quantum wells that have been partially depolarized due to the high density of injected charge. Either way, the numerator in Eq. (\ref{ee_def}) therefore varies by less than 1 \% and the observed $\Delta\textrm{V}_\textrm{F}$ in Fig. \ref{stats}(b) is indeed experimental proof of the charge injection hypothesis \cite{quan2014, li2016, ho2022, quevedo2024}.

\section{Conclusion}
The local $\textrm{V}_\textrm{F}$ reduction at V-defect rims reported in Fig. \ref{stats}(b) is close to the $\approx$ 1 V reduction observed in green-emitting LEDs containing V-defects \cite{lheureux2020} and as predicted in full, 3D calculations of their current-voltage characteristics \cite{ho2022}. This excellent agreement between the local measurements reported here and the macroscopic forward bias reductions, proves that in LEDs containing V-defects, the injected current passes essentially entirely through the multi-quantum well heterostructure via the $\left\{10\bar{1}1\right\}$ facets of the V-defects rather than through the $\left(0001\right)$-planes. Furthermore, the observation of small, 10 meV, blue-shifts in the electroluminescence emitted when the tip injects holes at the V-defect rims, indicates that the majority, if not all, of the injected holes radiatively recombine in the $\left(0001\right)$-plane quantum wells after lateral transport from the V-defect facets. The unique ability to simultaneously perform local electrical and optical spectroscopies at the nanoscale offered by the STLM reveals a clear picture of the role played by V-defects in WPE improvements recently obtained in long-wavelength-emitting nitride LED heterostructures.

\begin{acknowledgments}
Support was provided by the Simons Foundation (Grant No. 601952 for J.S.S. and No. 1027114 for C.F. and C.W. Additional support at UCSB was provided by the Solid State Lighting and Energy Electronics Center (SSLEEC); U.S. Department of Energy under the Office of Energy Efficiency \& Renewable Energy (EERE) Award No. DE-EE0009691; the Vannevar Bush Faculty Fellowship managed by the Office of Naval Research, Award No. N00014-25-1–2040; and the Korea Institute for Advanced Technology (KIAT) through Seoul Viosys Co., Ltd. (award SB250158). A portion of this work was performed in the UCSB Nanofabrication Facility, an open-access laboratory. ACHR and JP thank Didier Lenoir for the design and realization of the floating, differential current amplifier electronics.
\end{acknowledgments}

\appendix
\section{300 K characteristics of the commercial green-emitting LED}
\label{appendix_LED}

\begin{figure}
    \centering
    \includegraphics[width=0.8\columnwidth]{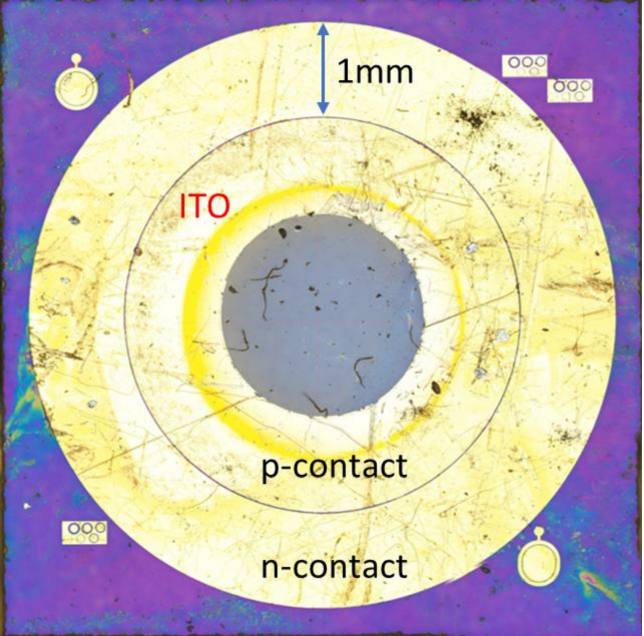}
    \caption{A photograph of the LED contact geometry taken from above. The concentric, circular geometry of the contacts is well adapted to studies with the STLM but results in current crowding at the inner edge of the contact. In devices that are not studied with the STLM, an ITO layer can be added, as shown, to avoid this.} 
    \label{photo_LED}
\end{figure}

The light-emitting diode (LED) studied here is a commercial green-emitting heterostructure manufactured by Seoul Viosys. The heterostructure is processed into contacted devices, a top view of which is shown in Fig. \ref{photo_LED}. This circular geometry is unusual for a commercial LED, but is adapted to studies with the scanning tunneling luminescence microscope (STLM). More specifically, the n-type and p-type Ohmic contacts to the LED's neutral zones are concentric circles around a central, circular emitting zone (diameter = 2.8 mm) where the STLM tip scans the surface.

\begin{figure}
    \centering
    \includegraphics[width=1\columnwidth]{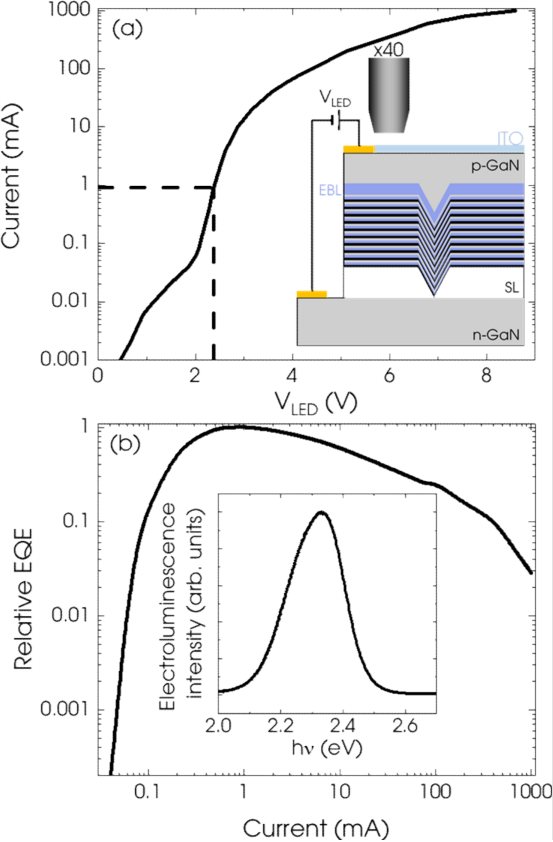}
    \caption{Macroscopic characteristics of the ITO-covered LED to be compared with those of the bare LED shown in the main article. (a) Current-voltage characteristics with a schematic cross-section of the LED heterostructure with V-defect shown inset. (b) The relative EQE and the electroluminescence spectrum measured at maximum EQE, inset.} 
    \label{macro_ito}
\end{figure}

In the samples used here, the circular emitting zone is terminated by the top p-GaN layers of the heterostructure. The macroscopic characteristics of the LED prior to removal of the top p-GaN layers (using atomic layer etching) for STLM studies, is shown in Fig. \ref{macro} where, as described, the measured electrical efficiency is 77 \%. This is somewhat lower than the $\approx$ 90 \% reported in state-of-the-art devices \cite{lheureux2020}. This is not because the heterostructures studied here are intrinsically poor, but is rather a result of the contact geometry and the details of the light collection. For example, in the circular geometry of Fig. \ref{photo_LED}, current crowding occurs at the inner circumference of the p-contact (when forward biased this is visually obvious as the LED is significantly brighter along this circumference). In an attempt to minimize current crowding, transparent (Indium Tin Oxide, or ITO) contacts are used. In the device shown in Fig. \ref{photo_LED} (which is not used in the STLM studies), a thin ($\approx$ 100 nm) ITO layer is deposited onto the circular emitting zone, and makes electrical contact with the p-contact of the LED (this is also shown in the cross-sectional sketch, inset in Fig. \ref{macro_ito}(a)). The edge of the ITO is visible in the photograph as a slightly darker yellow ring. Unfortunately, the addition of the ITO results in only minor changes to the current crowding so that a more representative estimation of the EE of the LED is still not possible.

Instead of relying on the ITO, the EE of the LED heterostructure is estimated using a local collection of the electroluminescence at the inner edge of the p-contact using a x40 long working distance microscope objective as indicated inset in Fig. \ref{macro_ito}(a). In this geometry light is collected only from the emissive part of the device. The resulting relative EQE curve is shown in Fig. \ref{macro_ito}(b), and peaks at an injected current of approximately 0.9 mA and a forward bias of V$_\textrm{LED}$ = 2.4 V as indicated by the dashed lines in Fig. \ref{macro_ito}(a). Using the electroluminescence spectrum at this forward bias shown inset in Fig. \ref{macro_ito}(b) where the emission centroid is approximately h$\nu$ = 2.3 eV, an electrical efficiency of 95 \% is estimated from Eq. (\ref{ee_def}). This is comparable to that reported in state-of-the-art green-emitting LEDs \cite{lheureux2020}.

\section{Scanning tunneling microscopy on the as-grown LED heterostructure}
\label{appendix_STLM}

\begin{figure}
    \centering
    \includegraphics[width=1\columnwidth]{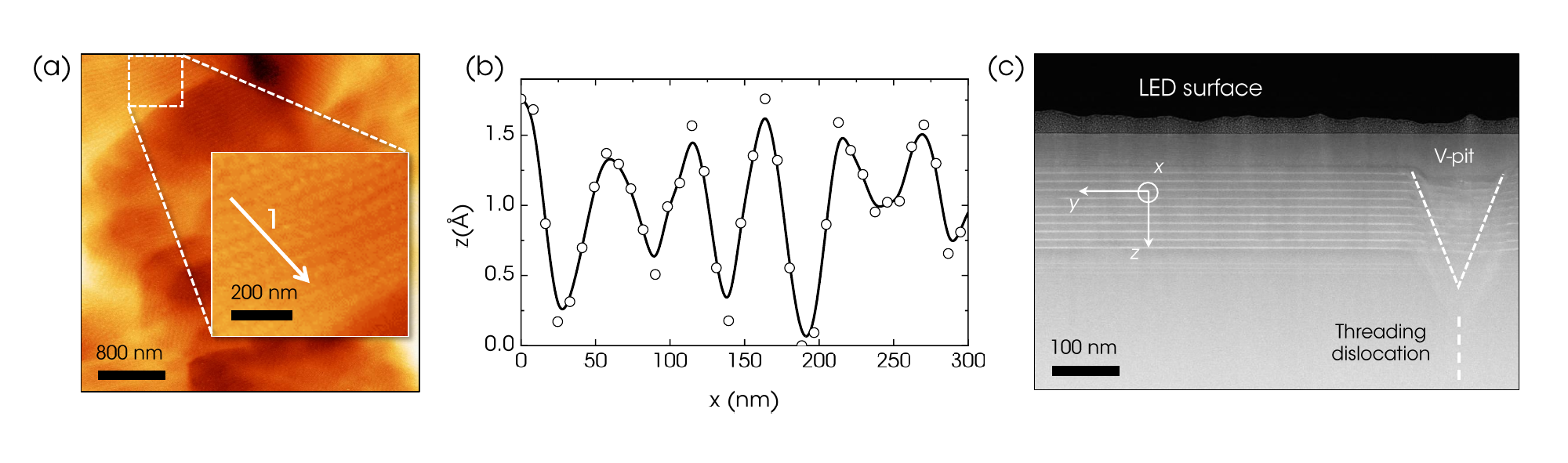}
    \caption{Topography of the operational LED with $\textrm{V}_\textrm{LED}$ = 0 prior to atomic layer etching. (a) The 300 K image obtained with $\textrm{V}_\textrm{TIP} = 5.5$ V and $\textrm{I}_\textrm{TIP} = 1$ nA. (b) A line profile of the surface along the line labelled 1 in (a). (c) A HAADF image of the heterostructure.} 
    \label{stm}
\end{figure}

Figure \ref{stm}(a) shows the scanning tunneling topography obtained at 300 K with $\textrm{V}_{\textrm{TIP}}$ = 5.5 V and a set-point tunnel current of $\textrm{I}_{\textrm{TIP}} =$ 1 nA. Currents at this intensity or lower are generally found to yield the best quality topographic images. The surface shows the typical atomic steps of GaN grown on a 0.2$^{\circ}$ miscut patterned sapphire substrate. This is clear in the line profile of Fig. \ref{stm}(b) where a characteristic step height of about 0.2 nm can be associated with half the lattice parameter of GaN along the $\left(0001\right)$-axis ($\approx$ 0.26 nm). However, no trace of V-defects are visible in the topography due to the lateral flow grow mode which fills them during the growth of the top, p-GaN layer. This filling of the V-defects is most obvious in the high-angle annular dark-field (HAADF) transmission electron microscope image of this heterostructure shown in Fig. \ref{stm}(c). The inability to identify V-defects at 300 K in the topography is the principle reason for the atomic layer etching which exposes them at the surface. The etched sample is the one on which all STLM data is presented in the main article.  

\section{Differential current measurements}
\label{appendix_diff}

\begin{figure*}
\includegraphics[width=2\columnwidth]{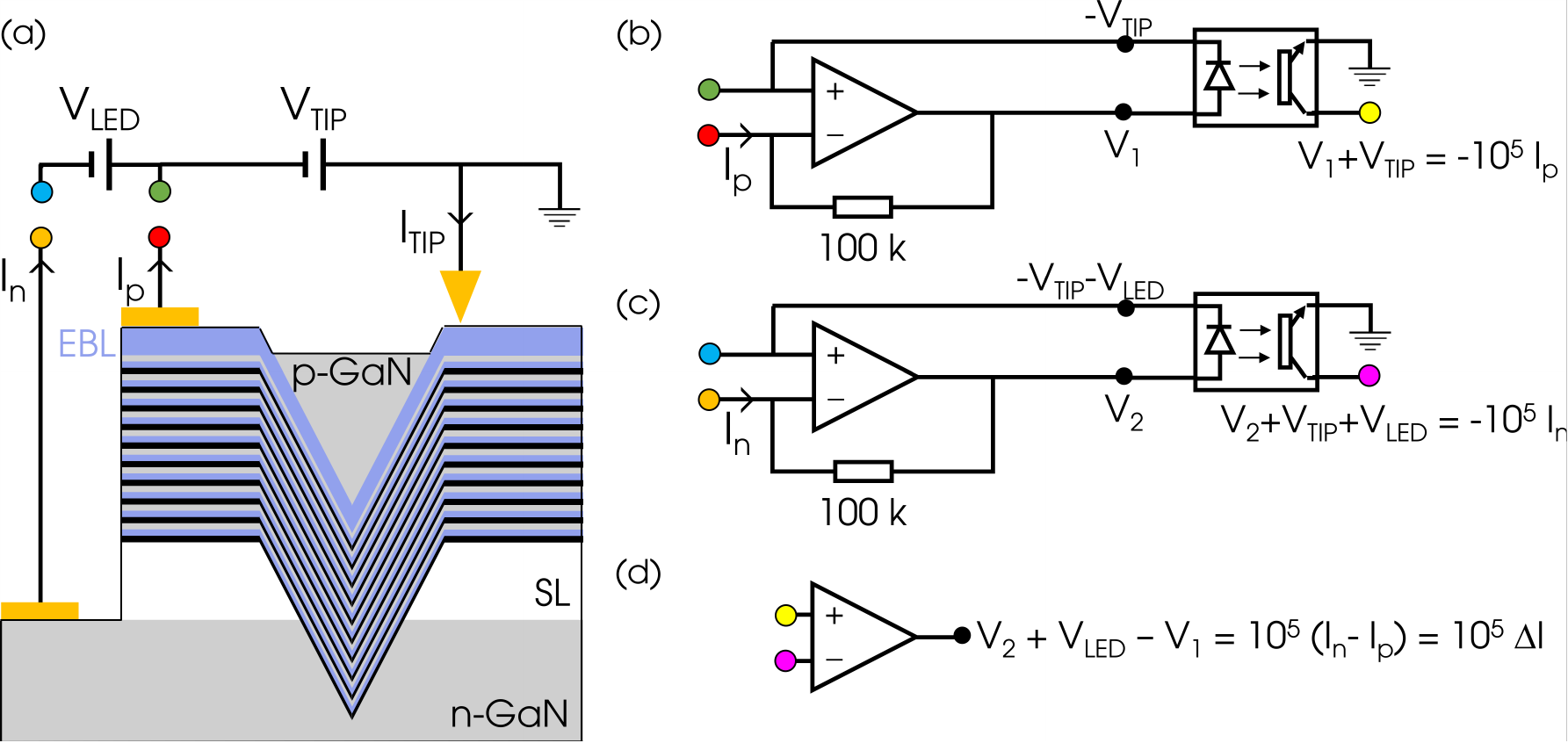}
\caption{\label{diff_current} The sample schematic as shown in the figure of the main article with details of the floating, differential current measurement shown. Floating current amplifiers shown in (b) and (c) are wired into the circuit in (a) according to the color-coded contacts. The outputs of these amplifiers are referenced to ground through an optocoupler and the resulting voltages (at the yellow and pink output contacts) are wired to the inputs of a unity gain differential voltage amplifier as shown in (d). The output voltage of this last amplifier is proportional to $\Delta I$, and this is mapped simultaneously with I$_{\textrm{TIP}}$, the topopgraphy, and the integrated electroluminescence intensity.}
\end{figure*}

Figure \ref{diff_current}(a) shows a schematic diagram of the atomic-layer etched LED heterostructure used for STLM studies. Details of the floating, differential current measurement are additionally shown. The two currents to be measured, $\textrm{I}_{\textrm{n}}$ flowing from the ohmic contact on the n-type neutral zone of the LED, and $\textrm{I}_{\textrm{p}}$ flowing from the Schottky contact on the electron blocking layer (EBL) are indicated. Each of these terminals is wired to the inverting input of an operational amplifier (OPA111) as shown in Fig. \ref{diff_current}(b) (red input wired to the p-Schottky contact on the LED) and Fig. \ref{diff_current}(c) (yellow input wired to the n-Ohmic contact on the LED). The non-inverting input on each of these operational amplifiers (green for the $\textrm{I}_{\textrm{p}}$ measurement, blue for the $\textrm{I}_{\textrm{n}}$ measurement) serves as the voltage reference in each case. As seen in Fig. \ref{diff_current}(a), this is ($-\textrm{V}_{\textrm{TIP}}$ in the case of the $\textrm{I}_{\textrm{p}}$ measurement, and $-\textrm{V}_{\textrm{TIP}}-\textrm{V}_{\textrm{LED}}$ in the case of the $\textrm{I}_{\textrm{n}}$) measurement. The feedback coupling the inverting input of each OPA111 to its output is made using a selectable gain resistor of value $\textrm{R}_{\textrm{G}}$. In the homemade circuits designed and built by us, $\textrm{R}_{\textrm{G}}$ can take one of the following values: 1 k$\Omega$, 10 k$\Omega$, 100 k$\Omega$, 1 M$\Omega$, 10 M$\Omega$, or 100 M$\Omega$. In the measurements reported here and in the main article $\textrm{R}_{\textrm{G}} = 100$ k$\Omega$.

\begin{figure*}
    \centering
    \includegraphics[width=2\columnwidth]{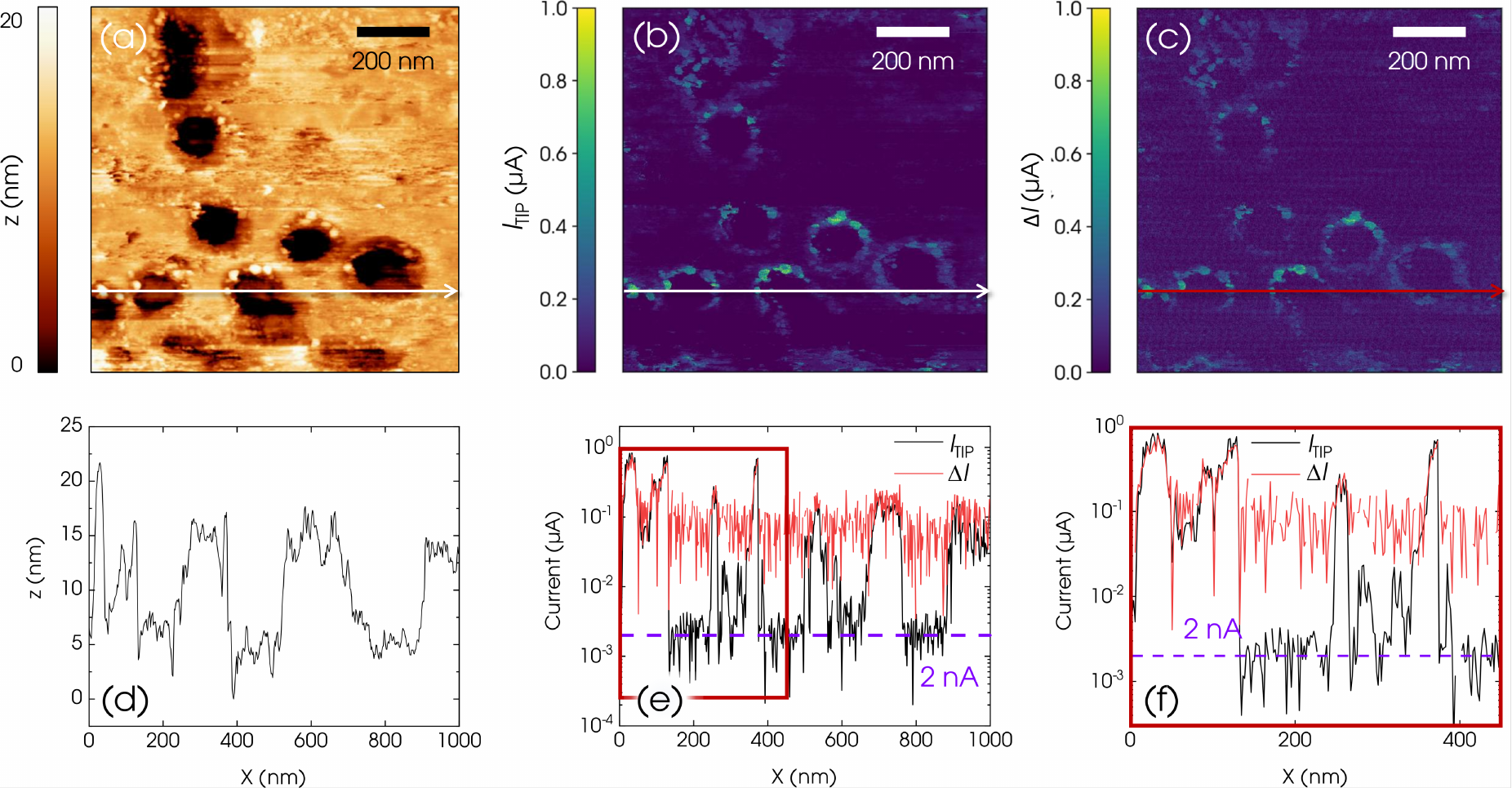}
    \caption{(a) (1) Effective STM topography of the ALE sample acquired at $\textrm{V}_\textrm{TIP} = 5.5$ V, $\textrm{I}_\textrm{TIP} = 2$ nA and $\textrm{V}_\textrm{LED} = 0$ V with the topography profile along the white line shown in (d). (b) Tip current map with the corresponding profile shown in black in (e). (c) Differential current map acquired with $\textrm{R}_\textrm{G} = 100$ $\textrm{k}\Omega$, with the line profile shown in red in (e). (f) A zoom of the current line profiles.} 
    \label{diff_current_measurements}
\end{figure*}

In the case of the $\textrm{I}_{\textrm{p}}$ measurement, the operational amplifier output voltage is $\textrm{V}_1$ which is referenced to $-\textrm{V}_{\textrm{TIP}}$, whereas for the $\textrm{I}_{\textrm{n}}$ measurement, the operational amplifier output voltage is $\textrm{V}_2$ referenced to $-\textrm{V}_{\textrm{TIP}}-\textrm{V}_{\textrm{LED}}$. In order to avoid the use of multiple levels of differential voltage amplifiers, the voltages are referenced to ground using optocouplers (IL300F) as shown in Fig. \ref{diff_current}(b) and (c). The resulting outputs are then: \begin{equation} \label{Ip} \textrm{V}_1 + \textrm{V}_\textrm{TIP} = -\textrm{R}_\textrm{G} \times \textrm{I}_\textrm{p} \end{equation} and \begin{equation} \label{In} \textrm{V}_2 + \textrm{V}_\textrm{TIP}+ \textrm{V}_\textrm{LED} = -\textrm{R}_\textrm{G} \times \textrm{I}_\textrm{n} \end{equation} respectively. Cabling of these outputs(yellow and pink) to the differential amplifier whose gain is 1 as shown in Fig. \ref{diff_current}(d), results in an an output voltage proportional to the desired current difference according to \begin{equation} \textrm{V}_\textrm{out} = \textrm{R}_\textrm{G} \times \left(\textrm{I}_\textrm{n} - \textrm{I}_\textrm{p}\right). \end{equation} 

Further details of the types of data obtained with the floating, differential current measurement are shown in Fig. \ref{diff_current_measurements}. Fig. \ref{diff_current_measurements}(a) shows the topography on part of the surface of the atomic layer etched sample where V-defects are clearly visible (see also main article). The topography is obtained with V$_{\textrm{TIP}}$ = 5.5 V, V$_{\textrm{LED}}$ = 0, and a set point current, I$_{\textrm{TIP}}$ = 2 nA. A corresponding line profile taken along the white line, is shown in (d). Along this particular line the height difference between the heterostructure's $\left(0001\right)$-plane and the remnant p-GaN filling the V-defects is approximately 10 nm. Typical values range from 10 nm to 20 nm (see also mask definitions in Fig. \ref{stats}. Fig. \ref{diff_current_measurements}(c) shows the I$_{\textrm{TIP}}$ map corresponding to the topography in Fig. \ref{diff_current_measurements}(a), with the large excess currents clearly visible at V-defect rims. A line profile of I$_{\textrm{TIP}}$ along the white line is shown as the black profile in Fig. \ref{diff_current_measurements}(e). 

The line profile of I$_{\textrm{TIP}}$ shows $\mu$A level currents at the V-defect rims, well above the set point current as discussed in the main article. When the tip scans above the remnant p-GaN filling the V-defects, the current feedback loop is able to re-establish the set point current (2 nA in this case). When injecting on the $\left(0001\right)$-plane, the feedback loop is also unable to stabilize the current at the set point, although excess currents tend to be smaller than those at the rims of V-defects. 

Fig. \ref{diff_current_measurements}(c) shows the $\Delta$I map measured using the floating, differential current amplifiers detailed in Fig. \ref{diff_current}. As discussed in the main manuscript, at the V-defect rims I$_{\textrm{TIP}} = \Delta$I, indicating that I$_{\textrm{n}} \gg \textrm{I}_{\textrm{p}}$ and that the tip-injected current is flowing through the heterostructure rather than leaking across the surface to the p-Schottky contact. This is also obvious by comparing the line profiles (black and red curves) in Fig. \ref{diff_current_measurements}(e). It is important to note that at the V-defect rims the passage of the current is accompanied by light emission at the wavelength of the quantum wells, further proof that the current is traversing the heterostructure.

Away from the V-defect rims it is seen that I$_{\textrm{TIP}} \neq \Delta$I, mainly because there is an offset in $\Delta$I of approximately 80 nA. This is clearly seen in the zoom of the current line profiles shown in Fig. \ref{diff_current_measurements}(f). The origin of this offset is unclear for the moment. It may either be an offset in one of the amplifiers shown in Fig. \ref{diff_current}, or it may be that V$_{\textrm{LED}}$ is not strictly speaking zero. In the latter case a small (80 nA) current will in fact circulate between the contacts of the LED without passing through the tip. Further tests are underway to identify the source of this current. It is reassuring to note that light is still emitted from the heterostructure at the wavelength of the quantum wells when the tip injects directly into the $\left(0001\right)$-plane between V-defects, although its intensity is lower than that measured at the V-defect rims where currents are higher as seen in Fig. \ref{EL}. It is only when the tip is injecting majority holes into the remnant p-GaN filling the V-defects that no light is emitted. In this case is it not yet possible to conclude whether the 2 nA set point current is leaking through the p-Schottky contact, or whether it is traversing the heterostructure assisted by non-radiative recombination at the threading dislocation at the center of the V-defect.

\section{Photoluminescence study}
\label{appendix_PL}

\begin{figure}
    \centering
    \includegraphics[width=1\columnwidth]{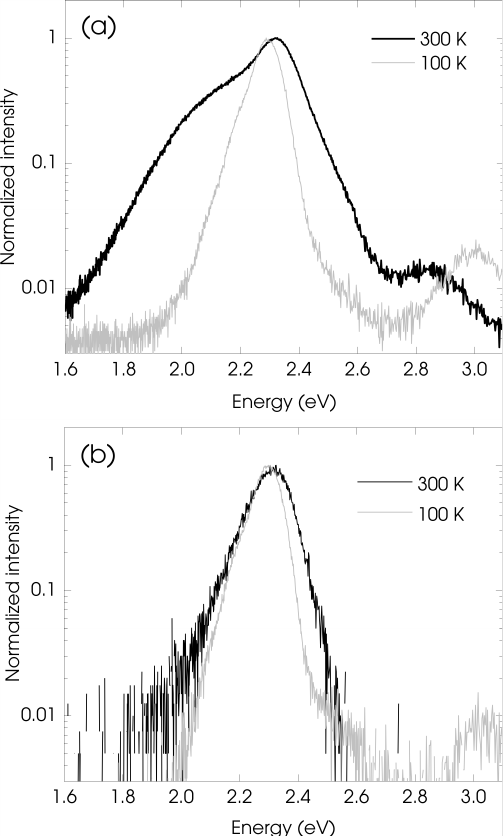}
    \caption{Photoluminescence of the operational short-circuited LED prior to atomic layer eatching at 300 and 100 K. The spectrum was obtained with a 3.3 eV laser excitation set at 8.9 µW and acquired with a 8 meV spectral resolution.} 
    \label{PL}
\end{figure}

Fig. \ref{PL} shows 300 K and 100 K photoluminescence spectra of the LED heterostructure before (Fig. \ref{PL}(a)) and after (Fig. \ref{PL}(b)) atomic layer etching. Spectra are obtained by pumping the central emitting zone of the LEDs with a laser tuned to 375 nm (3.3 eV). Prior to atomic layer etching of the p-GaN layer at the top of the heterostructure, the 300 K spectrum shows a low energy shoulder around 2 eV, and some high energy structure, particularly around 2.85 eV (see black curve, Fig. \ref{PL}(a)).

A clue to the identity of the low-energy peak can be found by looking at the 300 K photoluminescence spectrum of the atomic layer etched heterostructure (black curve, Fig. \ref{PL}(b)). The low energy shoulder is absent in this case. It is suggested therefore that the low-energy shoulder in Fig. \ref{PL}(a) arises from color centers in the p-GaN layers, for which there are a large number of possible candidates \cite{reshchikov2025}.

Also of interest is the temperature dependence of the photoluminescence. At 100 K in Fig. \ref{PL}(a), the high energy peak shifts to even higher energy ($\approx$ 3 eV) and becomes more intense. This peak, and the 300 K peak at 2.85 eV, are ascribed to the direct photoluminescence of the quantum wells aligned parallel to the V-defect facets \cite{yapparov2024}. As discussed in the main article, these quantum wells are thinner and of lower Indium content than the $\left(0001\right)$-oriented quantum wells, leading to bluer emission wavelengths. The semi-polar nature of the V-defect's $\left\{10\bar{1}1\right\}$ facets further contributes to the blue shift due to a relative absence of the quantum-confined Stark effect. The blue shift at lower temperatures from 2.85 eV to 3 eV is due to the bandgap dependence on the temperature, and the increase in intensity is due to a reduction in the diffusion length with temperature likely due to localization in the random potential resulting from compositional disorder in the $\left(0001\right)$-plane quantum well alloys i.e. lateral transport from of electrons and holes photo-injected at the V-defect facets to the LED's $\left(0001\right)$-plane is partially quenched at low temperature, allowing for a relative increase in radiative recombination in the quantum wells of the V-defect facets. This lateral transport of injected charge has been noted previously, but not under electrical injection conditions like those reported on in the main manuscript, and certainly not with the same spatial resolutions as those available with the STLM.

\bibliography{References}

\end{document}